\documentclass[aps,pra,showpacs,twocolumn]{revtex4}
\usepackage{graphicx}
\usepackage{amsmath}

\def\ket{\rangle}
\def\<{\langle}
\def\>{\rangle}

\begin{document}

\title{Modified Bennett-Brassard 1984 Quantum Key Distribution With Two-way
Classical Communications}
\author{Kai Wen$^{1}$ and Gui Lu Long$^{1,2}$ }
\affiliation{ $^1$ Key Laboratory For Quantum Information and
Measurements and Department of Physics, Tsinghua University,
Beijing 100084, China\\
$^2$ Key Laboratory for Atomic and Molecular NanoSciences,
Tsinghua University, Beijing 100084, China}
\date{\today }

\date{\today }

\begin{abstract}
 The quantum key distribution protocol without public announcement of
bases is equipped with a two-way classical communication symmetric
entanglement purification protocol. This modified key distribution
protocol is unconditionally secure and has a higher tolerable
error rate of 20\%, which is higher than previous scheme without
public announcement of bases.
\end{abstract}

 \maketitle

\section{INTRODUCTION}

 Quantum key distribution (QKD) is one of the most important and
exciting fields in quantum information. Its basic idea is to make
use of principles in quantum mechanics to detect whether there
exists an eavesdropper Eve, when two parties, Alice and Bob use
quantum channel to perform key distribution. In this way, the
security is much higher than that with only classical
communications. The earliest QKD protocol was proposed by Bennett
and Brassard in 1984 (BB84)\cite{BB84}. It is a kind of
prepare-and-measure QKD protocol, a protocol that Alice first
prepares a sequence of single photons, and she sends them to Bob
who measures each single photon immediately after receiving it.
Such kind of protocols are much more practical because they do not
require quantum computation and quantum memory.

 The security of QKD protocols is a basic problem in quantum
information. BB84 protocol has been proved to be secure when the
channels are noiseless. However, it is until recently that its
unconditional security  has been proved. Mayers\cite{Mayers} and
Biham\cite{Biham} presented their proofs, but the proofs are
rather complex. In Mayers' proof, BB84 protocol is secure when the
error rate of the channel is less than about 7\%.  Shor and
Preskill\cite{Shor} gave a much simpler proof which guarantees the
unconditional security of BB84 protocol if the error rate is less
than about 11\%. And then, Gottesman and Lo\cite{twoway} brought
two-way classical communications to BB84 protocol, and obtained a
much higher tolerable error rate, 18.9\%, which makes sure that
the BB84 protocol with two-way classical communications is
unconditional secure. Recently, Chau has presented a secure QKD
scheme making use of an adaptive privacy amplification procedure
with two-way classical communications whenever the bit error rate
is less than 20.0\%\cite{practical_twoway}.

 It is known that the standard BB84 protocol will use only half of
the transmitted qubits for key distribution. In order to enhance
the efficiency of the standard BB84 protocol, many variations have
been proposed. BB84 without public announcement of basis (PAB) is
just such a protocol\cite{pab}. In the eavesdropping detection
process of standard BB84, Alice announces her basis string in
which the qubit string is prepared, only after Bob has finished
receiving and measuring the qubit string. This announcement step
is called public announcement of basis(PAB). PAB guarantees Alice
and Bob to select the same measurement basis without
eavesdropping. However it also leads to waste of average one half
of the qubits. In BB84 without PAB, the communication parties do
not need PAB; instead, they agree on a secret random measurement
basis sequence before any steps of standard BB84. Alice encodes
qubits according to the prior basis sequence, and Bob uses the
same basis sequence to measure the qubits when he receives them.
In this way, none of the measurement results will be dropped as a
result of Alice and Bob choosing different measuring-basis.  BB84
without PAB, therefore, is still a prepare-and-measure QKD. In the
information processing of this protocol, Eve knows little about
the secret prior basis sequence yet, so all attacking strategies
that she can use are still the same as those in standard BB84. As
a result, the security of BB84 without PAB in noiseless channels
can be derived easily from the proof of the noiseless security of
standard BB84.

In this paper, we concentrate on the  security of BB84 without PAB
and its tolerable error rate. The protocol has been proved to be
secure through noisy channels following Shor and Preskill's
method\cite{pabshor} which obtains a tolerable error rate of 11\%,
the same as that of standard BB84\cite{Shor}. Recently, two-way
classical communications are introduced in security proof and it
increases the tolerable error rate of standard BB84 to
18.9\%\cite{twoway} and 20\%\cite{practical_twoway} respectively.
Inspired by this new idea, we prove the security of BB84 without
PAB with two-way classical communications. We first describe the
notations in this paper in section\ref{s2}. In section \ref{s3},
we present a QKD protocol without PAB and with a two-way
entanglement purification protocol(2-EPP), and prove its security.
Then we use a theorem in section \ref{s4} to reduce the protocol
into a prepare-and-measure protocol, that is, the BB84 without PAB
and with two-way classical communications, and gives a detailed
example in section \ref{s5} to obtain its minimal tolerable error
rate of 20\%. We give a brief summary in section \ref{s6}.

\section{Notations} \label{s2}

 The notations in this paper are mostly the same as those in  Gottesman and
Lo's paper\cite{twoway}.
 A Pauli operator acting on $n$ qubits is a $n$-dimension tensor product of
individual qubit operators that are of the following forms:
\begin{eqnarray}
 \notag  I(the\ identity), X=\left(\begin{array}{cc}\ 0 & \ 1 \\ \ 1 &
\ 0\end{array}\right), \\ Y=\left(\begin{array}{cc}\ 0 & -i \\
\ i & \ 0\end{array}\right), Z=\left(\begin{array}{cc}\ 1 & \ 0 \\
\ 0 & -1\end{array}\right).
\end{eqnarray}
Note that $X$, $Y$ and $Z$ operators are anti-commutative with
each other, and all the Pauli operators have only eigenvalues +1
and -1.

 Bell bases are the four maximally entangled states
\begin{equation}
\Psi^\pm = \frac{1}{\sqrt{2}} (|01\ket\pm|10\ket),\Phi^\pm =
\frac{1}{\sqrt{2}} (|00\ket\pm|11\ket).
\end{equation}

A \emph{symmetric EPP} can be described with a set of operators
$\{M_\mu\}$ plus unitary decoding operations $U_\mu \otimes (P_\mu
U_\mu)$, where $P_\mu$ is a Pauli operator. Each $M_\mu$ is a
particular measurement step of the protocol with the index $\mu$
which denotes a measurement history sequence in which each bit is
0 or 1 based on the outcome of the corresponding measurement step.
According to the history sequence $\mu$, Alice and Bob should both
choose the same operator $M_\mu$ to do measurement and this is why
the EPP is called symmetric. $U_\mu \otimes (P_\mu U_\mu)$ is
error correcting operations depending on $\mu$ after they obtain
all error syndromes. Alice performs $U_\mu$ while Bob performs
$P_\mu U_\mu$ operation.

 In various symmetric EPPs, there exist a set of the EPPs in which
all measurements $M_\mu$ are of eigenspaces of Pauli operators,
and the decoding operator $U_\mu$ is a Clifford group operator,
and the error-correcting operator $P_\mu$ is a Pauli operator.
This set of symmetric EPPs are called \emph{stabilizer EPPs}. And
if all measurements $M_\mu$ of a stabilizer EPP are either
$X$-type (including only I and X operators) or $Z$-type(including
only I and Z operators), and $U_\mu$ involves only CNOTs, this
stabilizer EPP is called a \emph{Calderbank-Shor-Steane-like
EPP}(CSS-like EPP). The CSS-like property comes from the idea of
CSS code which decouples the error-correction of X and Z and
guarantees the reduction in Shor and Preskill's proof of
BB84\cite{Shor}. Below, all the EPPs are CSS-like unless noted
explicitly. In CSS-code CSS($C_1,C_2$) is constructed from two
classical linear codes $C_1$ and $C_2$ that encodes $k_1$ bit and
$k_2$ bits of codewords into $n$-bits codewords, and $C_2\subset
C_1$ and $C_1$ and $C_2^\perp$ both correct $t$ errors.

 In EPP with one-way classical communications(1-EPP), Alice does
not know the measurement results of Bob and can not obtain the
history sequence $\mu$. Therefore, all the measurements and
operations in 1-EPP are independent to $\mu$. However in EPP with
two-way classical communications(2-EPP), Bob can also tell Alice
his measurement results through classical channels, so the
communication parties can make use of history sequence $\mu$ and
choose proper measurement operator according to the current
history, and the final decoding and error-correcting operations
also vary with the measurement results. In this way, 2-EPP are
supposed to tolerate higher error rate than 1-EPP, and we will
show that introducing two-way classical communications indeed
increases the tolerable error rate for BB84 without PAB.

\section{The QKD with 2-EPP without PAB}
\label{s3}

 In this section, we present a QKD protocol with 2-EPP without PAB,
and prove its security through noisy channels.

\emph{Protocol 1: QKD with 2-EPP without PAB}

\begin{enumerate}

\item Alice and Bob share a secret random $(2n/r)$ bit string, and
repeat it $r$ times to form a basis sequence $b$.

\item Alice prepares $2n$ EPR pairs in the state
$(\Phi^+)^{\otimes 2n}$, and applies a Hadamard transformation to
the second qubit of each EPR pair where the corresponding bit of
the basis sequence $b$ is 1.

\item Alice sends the second half of each EPR pairs to Bob.

\item Bob receives the qubits, and publicly announces the
reception.

\item Alice randomly chooses $n$ pairs of the $2n$ EPR pairs as
check bits to check the interference of Eve.

\item Alice broadcasts the positions of the check EPR pairs.

\item Bob applies a Hadamard transformation to the qubits where
the corresponding bit of the basis sequence $b$ is 1.

\item Alice and Bob both measure their own halves of the $n$ check
EPR pairs on the $Z$ basis, and publicly compare the results. If
there are too many disagreements, they abort the protocol.

\item Alice and Bob apply 2-EPP to the remaining $n$ EPR pairs,
and then share a state with high fidelity to $(\Phi^+)^m$.

\item Alice and Bob measure the state in the $Z$ basis to obtain a
shared secret key.
\end{enumerate}

 In protocol 1, the idea of QKD without PAB is applied in step 1, 2
and 7. Alice and Bob share a basis sequence $b$ at the beginning.
They can first distribute a smaller random sequence with bit
length $2n/r$ by another QKD protocol or other methods, then
repeat it $r$ times. Although the basis sequence $b$ is a repeat
of a random string, if $r$ is small and $n$ is large enough, the
information of the base sequence of Eve is still exponentially
small for $n$, and the effect of $r$ is only to increase the
information of the base sequence of Eve by multiplying polynomial
of $r$. Therefore, we can affirm that Eve knows very little about
the basis sequence.

 Knowing that Eve knows very little about $b$, we can follow the
method of Gottesman and Lo\cite{twoway} to derive the
unconditional security of protocol 1. First, protocol 1 is based
on stabilizer EPP, hence the quantum channel is equivalent to a
Pauli channel. Furthermore, because all operators in protocol 1
commute to each other, we can apply classical probability
analysis. Calculating the probability of the success of
error-correcting, because Eve knows little about the basis
sequence, we find that the fidelity of the state shared by Alice
and Bob after EPP to $(\Phi^+)^{\otimes m}$ is $1-2^{-s}$ for a
large factor $s$\cite{Shor}. By lemma 1 and lemma 2 in
\cite{LoChau}, Eve's mutual information with the final key is less
than $2^{-c}+ 2^{O(-2s)}$ where $c=s-\log _2 (2m+s+1/\ln 2)$. As a
result, Eve's information about the final key is exponentially
small and the unconditional security of protocol 1 is proved.

\section{BB84 with two-way classical communications without PAB}
\label{s4}

 Protocol 1 is based on EPP which requires quantum computers to
process. In this section, we will reduce protocol 1 to a
prepare-and-measure protocol, that is, BB84 with two-way classical
communications without PAB(2-BB84 without PAB). The equivalent
reduction of protocol 1 is based on the main theorem of Gottesman
and Lo\cite{twoway}. We revise it in a more simple way and apply
it to protocol 1 as the following.

\newtheorem{theorem}{Theorem}
\begin{theorem}[Revised Main Theorem in \cite{twoway}]
Suppose a 2-EPP is CSS-like and also satisfies the following
conditions:
\begin{enumerate}

\item If $M_\mu$ is $X$-type operators for a specific step with
$\mu$, for the following step with $\mu'$( $\mu'=\mu 0\ or\ \mu
1$), the choose of $M_{\mu'}$ is independent of the measurement
result of $M_\mu$, that is $M_{\mu 0}=M_{\mu 1}$.

\item The final decoding operations $U_\mu$ can depend arbitrarily
on the outcome of the measured $Z$-type operators, but cannot
depend on the outcomes of measured $X$-type operators at all. The
correction operation $P_\mu$ can depend on the outcome of $X$-type
operators, but only by factors of Z.
\end{enumerate}
Then protocol 1 can be converted to a prepare-and-measure QKD
without PAB scheme with the same security.
\end{theorem}

 The first condition in theorem 1 is equivalent to the tree diagram
representation of the first condition of the main theorem in
\cite{twoway}. If Alice and Bob drop the phase-error, they do not
know the exact result of the phase-error. In order to continue the
2-EPP, they must choose the unique measurement operator in the
next step despite of the $X$-type measurement outcome. The
existence of the two condition guarantees this statement. And a
CSS-like EPP makes that the corrections of bit-flip errors and
phase-flip errors are separated. So Alice and Bob can perform only
bit-flip error correction and do not require quantum computers to
correct phase-flip errors.

 In detail, the first step is to throw away $X$-basis operations
and measurements. The introduction of QKD without PAB in protocol
1 affects only in step 1 to 7, before the error correction and
privacy amplification. It only modifies the choice of base
sequence, and when the EPP is proceeded, all particles are in the
$Z$-basis. Therefore, the transformation of 2-EPP quantum circuit
in\cite{twoway} can be directly applied. After the transformation,
Alice and Bob can obtain a classical circuit with measurements
only in $Z$-basis.

 The second step is to transform the protocol into a
prepare-and-measure QKD. Following the same idea in \cite{Shor},
because all operations in protocol 1 commute with each other, it
is not necessary for Alice to prepare and distribute EPR pairs and
then measure them. Instead, Alice can measure them before
distribution. In other words, Alice can just prepare a random
binary string, and encode it into qubits, and send them to Bob.
Also, Bob can measure the qubits in the basis according to the
basis sequence immediately after he receives them, instead of
using quantum memory to store the qubits. Thus, we can
successfully transform protocol 1 into a prepare-and-measure QKD
without PAB.

 In the final step, in order to simplify the protocol, Alice and
Bob can perform 2-EPP to reduce the error rate of the qubits until
both bit-flip and phase-flip errors are lower than the bound of
the capacity of 1-EPP. Then they can perform 1-EPP to correct the
remaining error and obtain the final secret key\cite{twoway}.

 Consequently, we can conclude the content above into protocol 2 as
the following.

\emph{Protocol 2: Secure BB84 with two-way classical
communications without PAB}

\begin{enumerate}

\item Alice and Bob share a secret random $(2n/r)$ bit string, and
repeat it $r$ times to form a basis sequence $b$.

\item Alice prepares $2n$ random qubits, measure each qubit in
$Z$-basis which the corresponding bit of $b$ is 0 or in $X$-basis
which the corresponding bit of $b$ is 1. So Alice obtain an random
key and encode it in the qubit string.

\item Alice sends the qubit string to Bob.

\item Bob receives these $2n$ qubit string, measures it in
$Z$-basis or $X$-basis according to $b$, and then publicly
acknowledges the receipt.

\item Alice randomly chooses $n$ qubits as check bits and
announces their positions.

\item Alice and Bob compare the measurement results of the check
bits. if there are too many errors, they abort the protocol.

\item Alice and Bob use a classical circuit transformed from 2-EPP
to do error-correction until the error rates of both bit and phase
are lower than the bound of the capacity of BB84 with one-way
classical communications, for example $11\%$ in \cite{Shor}.

\item Alice and Bob use the method in BB84 with one-way classical
communications to perform final error-correction and privacy
amplification to obtain the key. For example, they can use CSS
Code to correct errors, and obtain the coset $\nu+C_2$ as the
secret key\cite{Shor}.

\end{enumerate}

 According to theorem 1, protocol 2 is equivalent to protocol 1.
Therefore protocol 2 is also unconditional secure through noisy
channels.

\section{An example of secure BB84 with two-way classical
communications and without PAB} \label{s5}

 In section \ref{s3}, we give the secure BB84 with two-way classical
communications without PAB, that is protocol 2. However, protocol
2 is still a theoretic scheme, and needs further study to exploit
its capacity. In this section, a particular 2-EPP from
\cite{twoway} is presented and transformed to the classical
circuit. We use this classical circuit in step 7 of protocol 2 so
that we can estimate the lower bound of the tolerable error rate
of protocol 2.

 Although the theorem 1 guarantees the security of protocol 2, it
is still necessary to find a practical 2-EPP that fulfills the
theorem's conditions. Such a 2-EPP is presented in \cite{twoway}
induced from the classical error-correction theory. This 2-EPP
contains alternating rounds of two major steps, that is, bit-flip
error-correction step(``B step'') and phase-flip
error-correction(``P step'') step:

 \textbf{B step}\cite{twoway}: Alice and Bob randomly permute all
the EPR pairs.Then they each measure their own local $Z \otimes Z$
in order to obtain the bit-flip error of the remaining output
pair. If the results of Alice and Bob are different, they estimate
that there is a bit flip on the remaining output pair, and discard
it. This step is similar to advantage distillation in classical
communications by Maurer\cite{Maurer}.

 \textbf{P step}\cite{twoway}: Alice and Bob randomly permute all
the EPR pairs. Then they group them into sets of three, both
measure $X_1 X_2$ and $X_1 X_3$ on each set. This step can be
transformed into a circuit that first perform a Hadamard
transformation on each qubit, two bilateral XORs, measurement of
the last two EPR pairs, and a final Hadamard transform. If Alice
and Bob disagree on one measurement, Bob estimates that the phase
error was probably on the first two EPR pairs and does nothing; if
both measurements disagree, Bob assumes the phase error was on the
third EPR pair and corrects it by performing a $Z$ gate. This step
is induced from 3-qubit phase-flip error correcting code, and will
reduce the phase-flip error rate if the error rate is low enough.

 The completed 2-EPP consists of alternating rounds of the two
steps above. In each round, Alice and Bob first perform a B step
to calculate bit-flip error syndromes. This is a $Z$-type
measurement step. Then Alice and Bob perform a P step to calculate
phase-flip error syndromes, which is a $X$-type measurement step.
And P step does not affect later operations. So the 2-EPP
satisfies the conditions of theorem 1. After P step, they estimate
the error rate of the qubits by sacrificing some of them to
measure. If the error rate is lower than the bound of BB84 with
one-way classical communication, that is, about 11\%, they use
Shor and Preskill's method to obtain final key\cite{Shor},
otherwise, they go on with another round of B and P steps.

 By transforming the circuit of the 2-EPP according to theorem 1,
we can get a more detailed protocol than protocol 2 as the
following.

\emph{Protocol 3: secure BB84 example with two-way classical
communications without PAB}

\begin{enumerate}

\item Alice and Bob share a secret random $(2n/r)$ bit string, and
repeat it $r$ times to form a basis sequence $b$.

\item Alice prepares $2n$ random qubits, measure each qubit in
$Z$-basis in which the corresponding bit of $b$ is 0 or in
$X$-basis in which the corresponding bit of $b$ is 1. So Alice
obtains an random key and encodes it in the qubit string.

\item Alice sends the qubit string to Bob.

\item Bob receives these $2n$ qubit string, measures it in
$Z$-basis or $X$-basis according to $b$, and then publicly
acknowledges the receipt.

\item Alice randomly chooses $n$ qubits as check bits and
announces their positions.

\item Alice and Bob compare the measurement results of the check
bits. If there are too many errors, they abort the protocol.

\item (B step)Alice and Bob randomly pair up their own bits. Alice
publicly announces the parity(XOR) of the values of each pair of
her own, that is, $x_{2i-1} \oplus x_{2i}$, and Bob also publicly
announces the parity of his corresponding pair, that is, $y_{2i-1}
\oplus y_{2i}$. If the parities agree, they keep one of the bits
of the pair. Otherwise, they discard the whole pair.

\item (P step)Alice and Bob randomly group the remaining bits in
to sets of three, and compute the parity of each set. They now
regard those parities as their effective new bits in later steps.

\item Alice and Bob sacrifice sufficient $m$ of the new bit pairs
to perform the refined data analysis publicly. They abort if the
error rate is too large. And if the error rate is low enough, they
go to the next step, otherwise, they return to step 7.

\item Alice and Bob randomly permute their pairs, and use Shor and
Preskill method\cite{Shor} with one-way classical communications
to perform final error-correction and privacy amplification. In
detail, it contains the following sub-steps:

\begin{enumerate}
\renewcommand{\labelenumii}{\arabic{enumii})}

\item Alice and Bob select a proper CSS($C_1$,$C_2)$ Code Q.

\item Alice randomly choose a codeword $u$ from classical linear
code $C_1$, and announces $u+v$, where $v$ is a remaining code
bits.

\item Bob subtracts $u+v$ from his code bits, $v+\epsilon$, and
obtains $u+\epsilon$, and then corrects it to a codeword $w$ in
$C_1$.

\item Because code $C_2$ in CSS Code Q is a subgroup of $F^n_2$
which is the binary vector space on $n$ bits\cite{goodqc}, and
$u-w \in C_2$, Alice and Bob use the coset of $u+C_2$ as the final
key.

\end{enumerate}

\end{enumerate}

Protocol 3 consists of detailed operations of each step, which can
be studied further, for example, the tolerable error rate.
Reviewing the discussion in this section, the introduction of QKD
without PAB does not affect the error-correction and privacy
amplification of protocol 3. Thus, we can estimate the tolerable
error rate of our protocol without PAB directly from the same
method in \cite{twoway} and \cite{practical_twoway}. Firstly, from
\cite{twoway}, in BB84, the 2-EPP by alternating B and P steps is
successful provided that the bit error rate is lower than 17.9\%.
Hence, protocol 3 is secure with the same upper bound of error
rate. However, Gottesman and Lo point out that alternating B and P
steps is not optimal, and based on other arrangements of such two
steps, the BB84 can achieve higher tolerable error rate of
18.9\%\cite{twoway}. Moreover, by applying adaptive privacy
amplification procedure with two-way classical communications in
Gottesman and Lo's method, Chau obtain that tolerable error rate
of BB84 scheme is 20.0\%\cite{practical_twoway}. Such
modifications in error correcting and privacy amplification
procedure can also be applied to our BB84 protocol with two-way
classical communications and without PAB. In conclusion, our
protocol is secure whenever the bit error rate is less than
20.0\%, which is higher than the result of BB84 with only one-way
classical communications and without Pab\cite{pabshor}.

\section{Conclusions and discussions}
\label{s6}

 In this paper, we have proved the unconditional security of simple
modification of standard BB84 protocol---BB84 without public
announcement of bases, by applying two-way classical
communications. In addition, we present a detailed protocol,
protocol 3, and follow other 2-EPP procedures\cite{twoway,
practical_twoway} to calculate a lower bound of the tolerable
error rate of the protocol. The result of about 20.0\%
demonstrates advantages of two-way classical communications over
one-way classical communications without PAB whose tolerable error
rate is about 11\%\cite{pabshor}. Compared to the previous BB84
protocol sets, this protocol benefits from both two-way classical
communications which tolerate higher error rate and the technique
without PAB which increases key generation rate. As a result, it
is much more efficient than previous protocols, and can be widely
used in future quantum communications.

\section*{Acknowledgment}
  This work is supported by the National Fundamental
Research Program Grant No. 001CB309308, China National Natural
Science Foundation Grant No. 10325521, 60433050, the Hang-Tian
Science Fund, and the SRFDP program of Education Ministry of
China.

\end{document}